\documentclass[sigconf]{acmart}

\usepackage{xcolor}
\newcommand{\Yes}[1]{\textcolor{green!60!black}{\textbf{Yes}$^{#1}$}}
\newcommand{\No}[1]{\textcolor{red}{\textbf{No}$^{#1}$}}
\newcommand{\NA}{\textcolor{gray}{N/A}}
\newcommand{\Partial}[1]{\textcolor{orange!80!black}{\textbf{Partial}$^{#1}$}}

\usepackage{tikz}

\usepackage{xcolor}
\usepackage{babel}
\usetikzlibrary{positioning, shapes, arrows}

\usepackage{graphicx}
\usepackage{textcomp}
\usepackage{xcolor}
\usepackage{comment}
\usepackage{tcolorbox}

\def\BibTeX{{\rm B\kern-.05em{\sc i\kern-.025em b}\kern-.08em
    T\kern-.1667em\lower.7ex\hbox{E}\kern-.125emX}}

\usepackage{url} 
\usepackage{comment} 
\usepackage{algorithm}
\usepackage{todonotes}
\usepackage{tikz}
\usepackage{xcolor}

\usepackage[utf8]{inputenc}

\usepackage{siunitx}
\usepackage{array}

\usepackage{makecell}
\usepackage{listings}
\usepackage{float}
\usepackage{balance}
\usepackage{hyperref}
\usepackage{xspace}
\usepackage{tcolorbox}
\usepackage{enumitem}
\usepackage{multirow}
\usepackage{pifont}
\usepackage{tabularx}
\usepackage[para,online,flushleft]{threeparttable}
\usepackage{colortbl}
\usepackage{subcaption}
\usepackage{adjustbox}
\usepackage{booktabs}
\usepackage[compatibility=false]{caption}
\usepackage{tabularx}
\usepackage{framed}

\usepackage{booktabs}
\usepackage[table]{xcolor}
\usepackage{tikz}
\usepackage{amsmath, amssymb} 
\usepackage[T1]{fontenc}      
\usepackage{lmodern}          

\ccsdesc{Software and its engineering}

\newboolean{showcomments}
\setboolean{showcomments}{true}
\ifthenelse{\boolean{showcomments}}
 { \newcommand{\mynote}[2]{
      \fbox{\bfseries\sffamily\scriptsize#1}
        {\small$\blacktriangleright$\textsf{\emph{#2}}$\blacktriangleleft$}}}
        { \newcommand{\mynote}[2]{}}

\usepackage{booktabs}
\usepackage[table]{xcolor}

\usepackage{algorithm}
\usepackage{algpseudocode}

\title{SIEVE: Towards Verifiable Certification for Code-datasets}

\author{Fatou Ndiaye MBODJI}
\affiliation{%
  \institution{University of Luxembourg}
  \country{Luxembourg}
  \city{Luxembourg}
}
\email{fatou.mbodji@uni.lu}

\author{El-hacen Diallo}
\affiliation{%
  \institution{University of Luxembourg}
  \country{Luxembourg}
  \city{Luxembourg}
}
\email{el-hacen.diallo@uni.lu}

\author{Jordan SAMHI}
\affiliation{%
  \institution{University of Luxembourg}
  \country{Luxembourg}
}
\email{jordan.samhi@uni.lu}

\author{Kui Liu}
\affiliation{%
  \institution{Huawei}
  \country{China}
}
\email{kui.liu@huawei.com}

\author{Jacques KLEIN}
\affiliation{%
  \institution{University of Luxembourg}
  \country{Luxembourg}
  \city{China}
  \city{Luxembourg}
}
\email{jacques.klein@uni.lu}

\author{Tegawendé F. BISSYANDE}
\affiliation{%
  \institution{University of Luxembourg}
  \country{Luxembourg}
  \city{Luxembourg}
}
\email{tegawende.bissyande@uni.lu}

\begin{document}



\begin{abstract}
Code agents and empirical software engineering rely on public code datasets, yet these datasets lack \emph{verifiable} quality guarantees. 
Static ``dataset cards`` inform, but they are neither auditable nor do they offer statistical guarantees, making it difficult to attest to dataset quality. Teams build isolated, ad-hoc cleaning pipelines. This fragments effort and raises cost. We present SIEVE, a community-driven framework. It turns per-property checks into \emph{Confidence Cards}—machine-readable, verifiable certificates with \emph{anytime-valid} statistical bounds. We outline a research plan to bring SIEVE to maturity, replacing narrative cards with \emph{anytime-verifiable} certification. This shift is expected to lower quality-assurance costs and increase trust in code-datasets.
\end{abstract}

\keywords{dataset certification, confidence sequences, code datasets, software engineering datasets, reproducibility, continuous auditing, community-driven validation, machine-verifiable evidence}

\maketitle

\section{Introduction}
Data underpins modern science and machine learning. It powers recommendation systems, code-generation tools, and products used at global scale. Yet dataset trust remains fragile: once published, we often cannot tell if a dataset is complete, clean, or legally compliant.  If a dataset contains biases or compliance failures, the flaws propagate, compromising research validity and seeding failures in deployed systems. Other domains (e.g., chip design, infrastructure) certify quality before use. However, for datasets, the foundation of empirical science, we still lack transparent, machine-verifiable certification.

Early documentation efforts set the norm for human-readable records: \emph{Datasheets for Datasets} formalized a structured questionnaire covering motivation, collection, and limitations~\cite{gebru2021datasheetsdatasets};
the \emph{Data Nutrition Label} proposed modular summaries to surface issues at a
glance~\cite{holland2018datasetnutritionlabelframework}; and \emph{Data Cards} emphasized
user-centric, purpose-driven documentation to aid responsible deployment~\cite{yang2024navigatingdatasetdocumentationsai}. 
To bridge prose and pipelines, recent work standardizes machine-readable metadata:
\emph{Open Datasheets} contributes a JSON schema to export structured
documentation that downstream systems can parse~\cite{roman2023open}; \emph{Croissant-RAI} define a Web-native
vocabulary for lifecycle, labeling, safety/fairness, and compliance, enabling direct load and validation of RAI (Responsible AI) metadata~\cite{jain2024standardized}. While these efforts standardize RAI integration, their effectiveness depends entirely on adoption by dataset providers.

In reality, dataset documents remain scarce. An audit of $7{,}433$ Hugging Face dataset cards found that only $30.9\%$ of repositories contain non-empty cards, although those datasets account for $95\%$ of downloads~\cite{geren2025blockchain}. Even among the most popular datasets, the critical section \emph{``Considerations for Using the Data''} which should describe biases, limitations, and downstream impacts averages only about $2.1\%$ of the content~\cite{geren2025blockchain}. At the same time, the \emph{EU AI Act} requires providers to publish training-data summaries and maintain technical documentation for regulatory oversight~\cite{eu-ai-act-2024}. The gap between regulatory expectations and current practice illustrates how far the ecosystem is from evidence-backed dataset certification.

Beyond under-documentation, risks are already materializing: widely adopted datasets may carry biases or violations, yet they have been used to support scientific conclusions. \cite{balloccu-etal-2024-leak} shows massive indirect leakage of benchmark data into closed-source LLMs during evaluation.

Code-datasets particularly differ from other corpora: they are executable artefacts whose auditing is both operationally and semantically demanding. In practice, audits require reconstructing toolchains, pinning compilers and package registries, resolving transitive dependencies, and running builds/tests whose outcomes can drift as ecosystems evolve. Meanwhile, repositories become inaccessible, APIs deprecate, new CVEs surface, and stale projects silently bias analyses—making “the same dataset” hard to reproduce across time and machines.

\begin{figure}[t]
\centering
\scriptsize
\resizebox{\columnwidth}{!}{%
\begin{tabular}{lccccc}
\toprule
\textbf{Property\footnotemark} &
\makecell{\textbf{CodeNet}\\ \scriptsize{\cite{puri2021codenet}}} &
\makecell{\textbf{CSNet}\\ \scriptsize{\cite{codesearchnet_repo}}} &
\makecell{\textbf{HumanEval}\\ \scriptsize{\cite{openai_humaneval_card}}} &
\makecell{\textbf{APPS}\\ \scriptsize{\cite{apps_card}}} &
\makecell{\textbf{The Stack v2}\\ \scriptsize{\cite{bigcode_stack_v2_card}}} \\
\midrule
Buildability     & \Yes{}     & \No{}     & \No{}     & \No{}     & \No{} \\
Test smoke       & \Partial{} & \Partial{}& \Partial{}    & \No{}     & \No{} \\
Link valid       & \No{}      & \No{}     & \NA        & \No{}     & \No{} \\
Dependency health& \No{}      & \No{}     & \No{}     & \No{}     & \Partial{} \\
License resolves & \No{}    & \Yes{}    & \No{}    & \No{}    & \Partial{} \\
\bottomrule
\end{tabular}}
\vspace{2mm} 
\begin{minipage}{0.95\columnwidth}
\scriptsize
\textbf{Evidence pointers:} \\
(Yes) \begin{itemize}
    \item CodeNet buildability in the "status" column;
    \item  CSNet: licenses for the source code in
the \_licenses.pkl 
\end{itemize}
 %
%
%
(Partial) \begin{itemize}
    \item CodeNet: tests provided but Only for AIZU;
    \item CSNet:  human relevance judgement are given;
    \item HumanEval: function to test generated in the "test" column;
    \item The Stack v2 : acknowledge that the training dataset could contain malicious code and the limitation of license attribution
\end{itemize} 
%
(N/A) \begin{itemize}
    \item HumanEval: APPS card: data are handwritten.
\end{itemize}

(No) \begin{itemize}
    \item Informations not found in the dataset cards
\end{itemize}

\end{minipage}
\caption{Documentation \emph{coverage} of practitioner--critical properties \emph{as advertised in dataset cards/docs}. 
\Yes{} = explicitly documented as addressed; \Partial{} = partially/indirectly stated; \No{} = not stated; N/A = not applicable.}
\Description{currently what is documlented in most used datsets.}
\label{fig:cardDiff}
\end{figure}


\footnotetext{Properties definitions: 
\texttt{buildability} = repo builds in a smoke run; 
\texttt{test\_smoke} = if tests exist, a short run passes; 
\texttt{link\_valid} = entries resolve to repo+commit; 
\texttt{dependency\_health} = vulnerable dependencies; 
\texttt{license\_resolves} = license present \& compatible.}

To better understand real needs, we conducted a survey (Cf. \ref{sec:interview}) from which we identified recurring properties required by code datasets. Figure \ref{fig:cardDiff} contrasts what popular code-ataset cards currently document with these needs.

\begin{tcolorbox}[colback=gray!5!white,colframe=gray!75!black,title= Gap in Datasets and Objectives, 
                  boxrule=0.8pt,
                  arc=2mm, 
                  left=4pt, right=4pt, top=4pt, bottom=4pt] 

\textbf{Gap.} While the ecosystem is converging on standards for where information should reside (Croissant-RAI), and regulators are demanding more (EU AI Act), to the best of our knowledge, there is no measurable evidence on the quality of code datasets, and even less concerning the properties demanded by researchers and practitioners.

\textbf{Objective.} SIEVE: the pioneering solution toward a transparent, machine-verifiable, per-property certificate for code datasets, reporting quality with anytime-valid statistical bounds. These certificates provide verifiable proof of dataset quality.

\end{tcolorbox}

\newcommand{\ts}[1]{\textsuperscript{#1}}

\section{ Understanding Dataset Challenges}

\label{sec:interview}

This section investigates practical challenges encountered when using code datasets.  
\subsection*{Interview}
We conducted semi-structured \cite{hove2005experiences} interviews. The details are given in the table \ref{tab:interview_methodology}.

\begin{table}[h!]
\centering
\begin{tabular}{p{2cm}p{6cm}}
\hline
Recruitment & Participants contacted with study overview \\
\hline
Format &  online or face-to-face \\
\hline
Participants & 18: (15 SE researchers and 3 AI engineers)\\
\hline
Focus & Dataset quality challenges \\
\hline
\end{tabular}
\caption{Interview methodology summary}
\label{tab:interview_methodology}
\end{table}

\begin{table}[h!]
\centering
\resizebox{\columnwidth}{!}{%
\begin{tabular}{@{}cl@{}}
\toprule
\# & Interview Question \\
\midrule
1 & What common quality challenges have you encountered in code datasets? \\
2 & How have you identified concerns or issues in datasets you worked with? \\
3 & What suggestions do you have for improving dataset documentation and reporting of issues? \\
4 & Can you provide examples of specific datasets where such issues were observed? \\
\bottomrule
\end{tabular}%
}
\caption{Key questions asked during the interviews.}
\label{tab:interview_questions}
\end{table}

\textbf{Key Findings:}

\begin{table}[h!]
\centering
\footnotesize
\caption{Key interview insights on code–dataset issues}
\label{tab:interview_insights}
\begin{tabular}{p{2.6cm} p{5.2cm}}
\toprule
\textbf{Aspect} & \textbf{Observation} \\
\midrule
Indirect discovery (compliance) & Compliance risks (licensing) are rarely detected directly; they surface via colleagues, talks, or reviews. \\
\midrule
Missed or low‑quality capture & Valuable data is often not captured; indiscriminate scraping and weak filters yield noisy or low‑quality corpora. \\
\midrule
Abandonment pattern & Teams frequently invest time, then abandon datasets due to quality issues; many cannot later recall the dataset names. \\
\midrule
Recall shaped by feedback & Datasets criticized by reviewers or reused by peers are more salient than those abandoned quietly. \\
\bottomrule
\end{tabular}
\end{table}

As summarized in table~\ref{tab:interview_insights}, our interviews with SE researchers and practitioners surfaced three recurring patterns: (i) dataset issues are \emph{rarely reported} and projects are often \emph{quietly abandoned}, wasting effort; (ii) \emph{compliance and policy risks} (e.g., licensing, sensitive content) are typically discovered \emph{indirectly and late} in the workflow; and (iii) even within the same group, teams in different SE subareas \emph{do not share signals}, so common risks remain invisible. In short, quality problems are discovered \emph{reactively} rather than \emph{proactively}. These observations motivate our approach: replace ad‑hoc, one‑off cleaning with a \emph{proactive,  certification layer}. Accordingly, we are designing a \emph{systematic analysis} of widely used code datasets to identify concrete manifestations of these issues and to prioritize the property definitions and pinned oracles that SIEVE will certify.

Informed by the insights from these interviews and targeting a potential solution, below, we present our proposal: SIEVE. 

\section{Proposed Framework: SIEVE}

As datasets gain value, public and private stakeholders invest heavily in cleaning and maintaining ever-changing corpora. They need continuous, reproducible assurance of quality, yet current efforts are fragmented and often duplicate the same dataset pre-process work. SIEVE empowers the stakeholder consortium to co-sponsor datasets and collaboratively refine their quality and properties on an ongoing basis. It also transforms checks into transparent, machine-verifiable certificates with quantitative guarantees, thereby reducing redundant effort and enhancing trust.

\subsection{Global View}

\begin{figure*}[h!]
  \centering
  \begin{subfigure}[c]{0.3\linewidth}
    \centering
    \includegraphics[width=\linewidth]{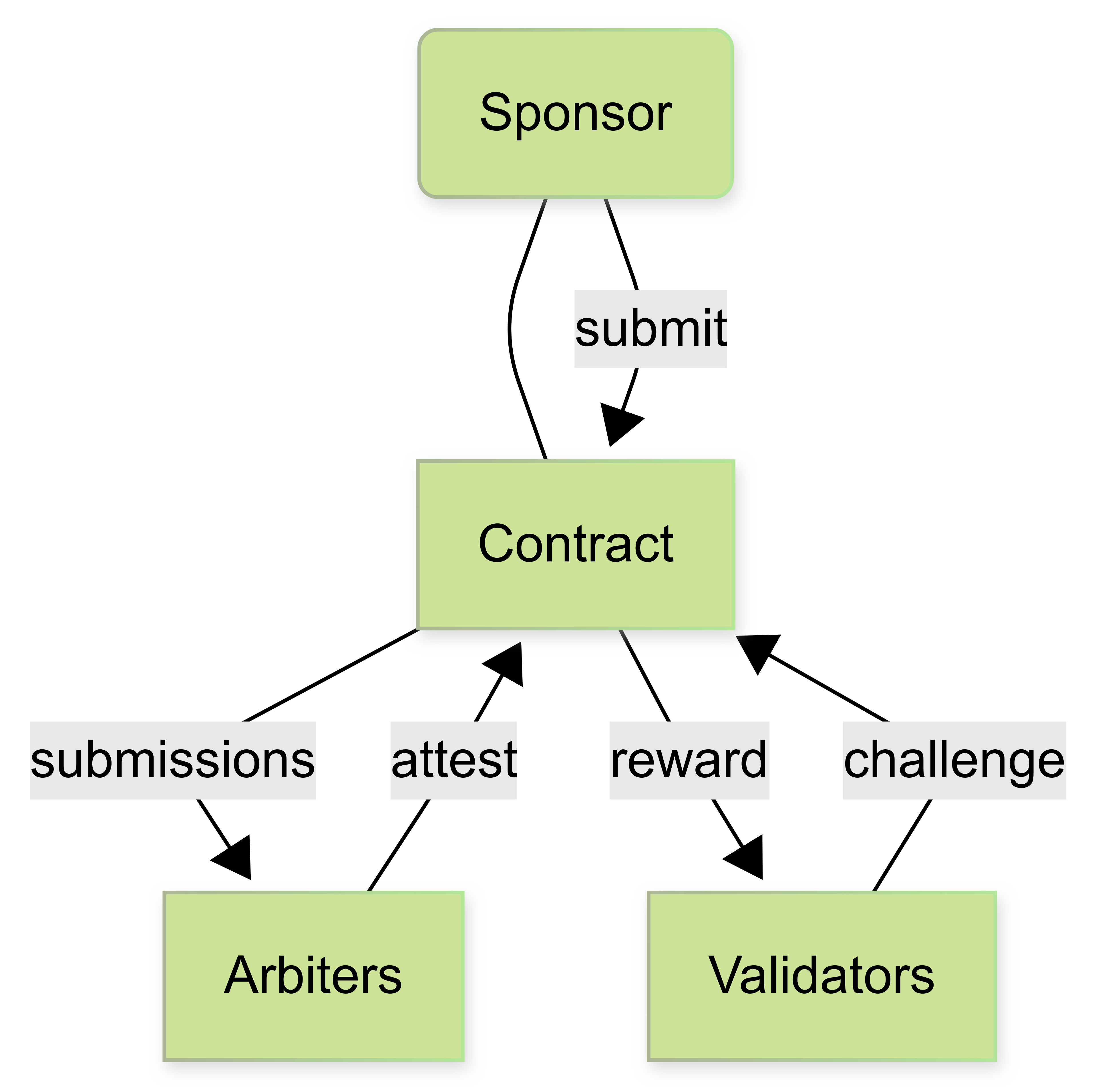}
    \Description{Overview of SIEVE}
    \captionsetup{width=\linewidth}
    \caption{}
    \label{fig:sieve}
  \end{subfigure}
  \hfill
  \begin{subfigure}[c]{0.6\linewidth}
    \centering
    \includegraphics[width=\linewidth]{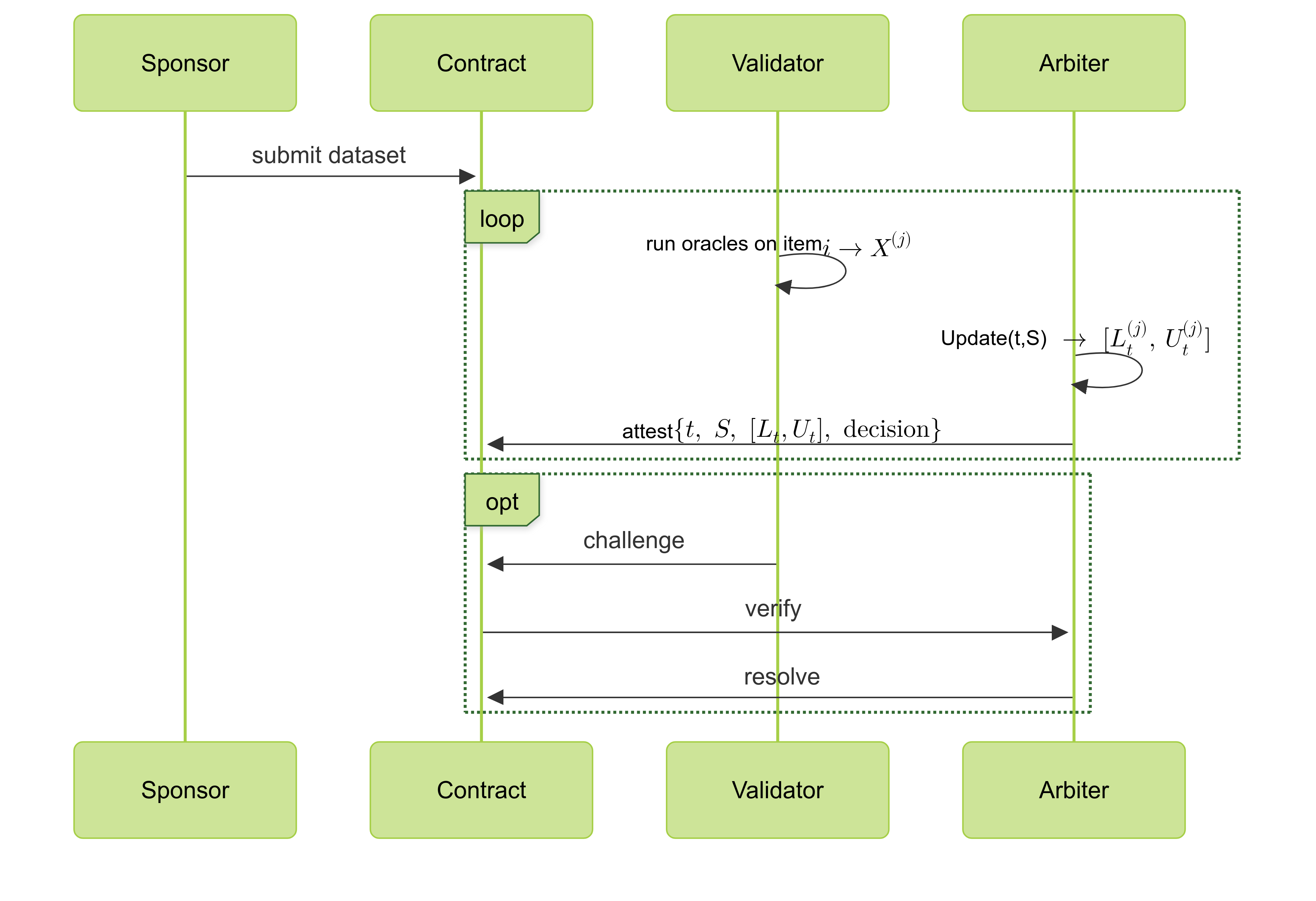}
    \captionsetup{width=\linewidth}
    \caption{}
    \label{fig:seq}
  \end{subfigure}
  \caption{Overview of SIEVE: (a) Global view, (b) Workflow.}
  \label{fig:sieve_overview}
\end{figure*}

\subsubsection*{Actors and roles} 
As depicted in Fig.\ref{fig:sieve}, \textbf{sponsors} submit datasets for audit. They also bear the cost of processing the entire audit and provide rewards as incentives for validators. The reward is assumed to be a recognition asset, similar to academic contributions such as reviewing papers. In scenarios involving private entities, continuous submission of their local test datasets for validation may be directly enforced by the sponsors. Sponsorship-related business models fall outside the scope of this work.

Because reviewing datasets containing hundreds of millions of records is both complex and expensive, sponsors may not require a full row-by-row assessment. Therefore, we introduce two tolerance measures per property: (i) an error bound $\varepsilon$, which specifies the accepted error on a given property, and (ii) a coverage parameter $(1-\delta)$, which limits the cost derived from auditing.

\textbf{Validators} are dataset users (e.g., researchers, engineers) who derive the public samples and run lightweight property checks (\texttt{oracles}) on these samples. Through sponsors, validators may also define properties aligned with their needs.

\textbf{Arbiters} reproduce validator evidence, aggregate results, and \emph{attest} the current confidence score. Their role can be configured differently depending on the deployment. In an academic context, arbiters may act as reviewers who simply aggregate and recheck validators’ claims; in other settings, AI models could serve this role. In all cases, arbiters are auditable, and validators may challenge their outputs. If a conflict arises, a contradiction report is issued to highlight violations of the attestation.
 
 \subsubsection*{Smart Contract} SIEVE leverages a contract \footnote{https://ethereum.org/smart-contracts/; accessed on \today{}} as a trust anchor that makes a dataset audit transparent and verifiable for stakeholders. It anchors the dataset and audit rules, fixes public randomness for unbiased sampling, escrows and settles funds under transparent rules, and keeps an append-only log of attestations and challenges. All checks run as off-chain evidences; the chain stores only commitments, ensuring independence from the sponsor and reproducible audits with an on-chain footprint.

\subsection{Confidence Card}
\label{sec:cs}


A \emph{Confidence Card} is a machine-readable record stating, for a dataset version and a
binary property $P$ (violation/no-violation), the current evidence:
sample count $t$, observed violations $S_t$, a live interval $[L_t,U_t]$ for the true
violation rate $p$, and a decision state. It is updated as more items are checked and can be replayed by any third party.
We use anytime-valid confidence sequences (CS): at every sample count $t$ (number of
distinct items evaluated), CS provide an interval for $p$ that remains valid no matter
when we look or stop (continuous monitoring).

\noindent\textbf{Assumptions.}
Uniform seeded sampling. deterministic, version-pinned oracle; tolerance $\varepsilon$ and coverage $1-\delta$ fixed.

\noindent\textbf{Guarantee.}
We maintain $[L_t,U_t]$ such that
\[
\Pr\!\big(\forall t\ge1:\ p\in[L_t,U_t]\big)\ \ge\ 1-\delta,
\]
valid under arbitrary peeking/stopping.

\noindent\textbf{Construction (Bernoulli, KL time-uniform).}
Let 
\[
d(a \,\|\, b) \;=\; a \log\!\left(\frac{a}{b}\right) + (1-a) \log\!\left(\frac{1-a}{\,1-b}\right),
\]
denote the binary Kullback--Leibler divergence between Bernoulli parameters $a$ and $b$, and define the anytime penalty
\[
\psi_t(\delta) \;=\; \log\!\left( \frac{2 \log_2(2t)}{\delta} \right),
\]
as in~\cite{howard2021timeuniform,ville1939,grunwald2019safetest}.

At each time $t$, we invoke a standard routine that maps $(t, \widehat p_t=S_t/t, \delta)$ 
to a confidence interval $[L_t, U_t]$ using a time-uniform Bernoulli bound 
(we adopt the KL-based formulation of~\cite{howard2021timeuniform}). Specifically,
\[
\begin{cases}
U_t = \inf \Bigl\{\, u \in [\widehat p_t, 1] :
      t \, d\!\left(\widehat p_t \,\|\, u\right) \;\ge\; \psi_t(\delta) \Bigr\}, \\[1ex]
L_t = \sup \Bigl\{\, \ell \in [0, \widehat p_t] :
      t \, d\!\left(\widehat p_t \,\|\, \ell\right) \;\ge\; \psi_t(\delta) \Bigr\}.
\end{cases}
\]

\subsection{Workflow}
This section presents the SIEVE workflow, which is structured into the following steps and illustrated in Fig. \ref{fig:seq}:

\begin{enumerate}
    \item  The sponsor submits a dataset for audit including:
\begin{itemize}\itemsep2pt
  \item \texttt{DatasetID} $=(\texttt{rootHash},\texttt{URLs})$ : exact dataset version (e.g., commit SHA/CID) and eventual link to the dataset.
  \item \texttt{Property set} $\mathcal{P}=\{(P_j,\varepsilon_j,\delta_j)\}_{j=1}^J$ 
  
  \item \texttt{Oracles}: content digests (e.g., repo+commit) of the checker for each $P_j$.
\end{itemize}

 \item The contract rejects duplicates for the same \texttt{rootHash} and locks a public randomness seed.
All parties derive the same uniform schedule of indices via a pseudorandom function.

\item  Repeated until a terminal decision:
\begin{enumerate}\itemsep2pt
  \item  Validators submit the next unclaimed seeded index; arbiters enforce membership and de‑dup.
  \item  For each property $P_j$, compute $X^{(j)}\!\in\!\{0,1\}$ on the sampled item.
  \item Publish \{\texttt{indices}, \texttt{bits}, \texttt{oracles}, \texttt{logs}\} to a off-chain store (e.g., IPFS) and its digest/URI on-chain.
  \item Arbiters reproduce the pack, update $(t,S_t^{(j)})$, and call
        to obtain $[L_t^{(j)},U_t^{(j)}]$ for each $P_j$ (Sec.~\ref{sec:cs}).
  \item  Arbiters co‑sign \(\texttt{attest}(t,S,[L_t,U_t],\texttt{decision})\);
  \item \textit{Stopping rule.} 
\[
\text{State}(t) \;=\;
\begin{cases}
  \textsc{Clean}, & \text{if } \forall j \;:\; U_t^{(j)} \le \varepsilon_j, \\[6pt]
  \textsc{Dirty}, & \text{if } \exists j \;:\; L_t^{(j)} \ge \varepsilon_j, \\[6pt]
  \textsc{Pending}, & \text{otherwise.}
\end{cases}
\]

\end{enumerate}

\item When a terminal decision is reached, the per‑property card is stored by content address and referenced on‑chain next to
\texttt{rootHash} and \texttt{seed}.

\begin{tcolorbox}[title=SIEVE Confidence Card, colback=gray!5!white,colframe=gray!75!black, boxrule=0.4pt]
\small\[
\small
\begin{array}{ll}
\texttt{dataset}:   & \{\texttt{rootHash},\, \texttt{seed}\} \\[2pt]
\texttt{property}:  & (P_j,\, \varepsilon_j,\, \delta_j,\, \texttt{oracle\_digest}) \\[2pt]
\texttt{evidence}:  & \big(t,\, S_t^{(j)},\, \widehat p_t^{(j)}{=}S_t^{(j)}/t,\, [L_t^{(j)}, U_t^{(j)}]\big) \\[2pt]
\texttt{decision}:  & \big(\text{State}(t),\ \mathrm{T2}\varepsilon_j\ \text{ if State}(t)=\textsc{CLEAN}\big)
\end{array}
\]

\textbf{Reading rule.}\quad
\textsc{CLEAN} iff \(U_t^{(j)} \le \varepsilon_j\);\quad
\textsc{DIRTY} iff \(L_t^{(j)} \ge \varepsilon_j\);\quad
otherwise \textsc{PENDING}.\\
\emph{\underline{Cleanliness lower bound}}: \(1 - U_t^{(j)}\) at coverage \(1-\delta_j\).

\textbf{Example.}\;
Let \(P_j=\texttt{buildability}\), \(\varepsilon_j=0.5\%\), \(1-\delta_j=95\%\).
Suppose the card shows \(t=2{,}500\), \(S_t^{(j)}=7\Rightarrow \widehat p_t^{(j)}=0.28\%\),
and \([L_t^{(j)},U_t^{(j)}]=[0.13\%,\,0.48\%]\).
Since \(U_t^{(j)}=0.48\%\le 0.5\%\), the decision is \textsc{CLEAN} and
\(\mathrm{T2}\varepsilon_j=2{,}500\). The dataset’s certified cleanliness for this property is
at least \(1-U_t^{(j)}=99.52\%\) (with 95\% anytime coverage).
\end{tcolorbox}

\item Validators may challenge arbiters \texttt{challenge(auditId, t, evidence\_uri)}. The contract records the resolution (and any penalties in incentive-enabled deployments).

\end{enumerate}

By aligning sponsors needs for clear guarantees with an efficient community participation, SIEVE turns ad-hoc, duplicated preprocessing into a transparent, replayable audit. Each dataset version receives a machine-readable \emph{Confidence Card} that (i) states what was checked and with what tolerance, (ii) publishes live, anytime-valid bounds, and (iii) is tamper-resistant (pinned oracles, reproducible sampling, content-addressed records). Thus, we bring less duplicated cleaning (shared, reusable evidence), lower onboarding cost for downstream users (i.e., cards become portable to CI/catalogs), and higher trust for all stakeholders (decisions are auditable and hard to game), without full rescans of the whole dataset.

\section{Future Plans}
Our future plans focus on operationalizing \textsc{SIEVE} beyond the core statistics (Sec.~\ref{sec:cs}) so that (RQ1) evidence is captured with near-zero friction inside developer tools, (RQ2) individual cleaning effort and duplication measurably decrease, and (RQ3) the framework demonstrably delivers value in real-world settings.

\subsection{Editor/CI integration (RQ1):} Ship a lightweight \textbf{SIEVE-Client} (VS Code/JetBrains)  that opportunistically captures build/test/dependency signals, packages an \emph{EvidencePack} with one-click consent, and submits it.  

\subsection{Efficiency \& cost (RQ2):} Add cache/skip rules for heavy checks, artifact/layer reuse, and a dashboard that tracks sample efficiency (T2$\varepsilon$), cleanliness growth $(1-U_t)$, and cost per certified point.  

\subsection{Deployment (RQ3):} Run multi-dataset pilots, publish public cards/artefacts (\texttt{rootHash}, \texttt{seed}, oracle, evidences), and wire cards to data catalogs.\\

Following this plan we expect, reproducible pipeline where editors/CI make evidence “nearly free”, cards certify properties with anytime-valid bounds, and pilots show measurable reductions in duplicated cleaning effort and increased trust thus validating the SIEVE for community-driven, per-property dataset certification.

\section{Conclusion}

We introduced SIEVE,  a community‑driven framework   that turns dataset-quality claims into anytime-valid statistical certificates. without scanning entire datasets. Our goal is to make SIEVE a lightweight yet dependable layer: a card schema, a library, pinned oracles for common properties, and easy editor/CI clients. Dataset hubs and CI systems can consume cards to enforce gates or display cleanliness lower bounds. Practitioners stop rebuilding private filters; instead, they contribute evidence that improves a shared, \emph{anytime‑verifiable} certificate. 

\section{Acknowledgments}
This work is supported by the Luxembourg Ministry of Foreign and European Affairs through their Digital4Development (D4D) portfolio under the project LuxWAyS (Luxembourg/West-Africa Lab for Higher Education
Capacity Building in CyberSecurity and Emerging Topics in ICT4Dev.)

\end{document}